\documentclass[twocolumn,showpacs,preprintnumbers,amsmath,amssymb]{revtex4}

\usepackage{amssymb, amsmath}
\usepackage[pdftex]{graphicx}
\usepackage[pdftex]{hyperref}

\begin{document}

\author{Steven Johnston}
\affiliation{Theoretical Physics, Blackett Laboratory, Imperial College London, South Kensington, London, SW7 2AZ, UK.}
\email{steven.johnston02@imperial.ac.uk}

\title{Feynman Propagator for a Free Scalar Field on a Causal Set}

\newcommand{\be}{\begin{equation}}
\newcommand{\ee}{\end{equation}}
\def\p{\hat{\phi}}
\def\lv{\langle 0 |}
\def\rv{| 0 \rangle}
\def\adag{\hat{a}^\dagger}
\def\a{\hat{a}}
\def\CS{\mathcal{C}}
\def\link{\prec\!\!*\,}
\def\Ale{\bar{A}}
\def\mink{\mathbb{M}}
\def\cint{\mathbb{A}}
\def\X{X}
\def\Y{Y}

\begin{abstract}
The Feynman propagator for a free bosonic scalar field on the discrete spacetime of a causal set is presented. The formalism includes scalar field operators and a vacuum state which define a scalar quantum field theory on a causal set. This work can be viewed as a novel regularisation of quantum field theory based on a Lorentz invariant discretisation of spacetime.
\end{abstract}


\maketitle

The possibility that spacetime is fundamentally discrete presents one
way to regulate the divergences in quantum field theory and the singularities in general relativity.
Causal set theory provides a model of discrete spacetime in which spacetime
events are represented by elements of a causal set---a locally finite, partially
ordered set in which the partial order represents the causal relationships between events.

Causal sets can be considered as physical spacetimes in their own right \cite{CausetMyrheim} (in particular, in attempting
to solve the problem of quantum gravity \cite{CausetHooft, Causet1, Causet2}) or they can be considered as merely Lorentz invariant discretisations of
continuum spacetimes: Lorentzian random lattices. Here we develop a formalism for quantum field theory on a causal set which is applicable to both points of view (alternative approaches include \cite{RS, SB}). The main result is the construction of the Feynman propagator for a real scalar field on a causal set. This work extends previous results for propagators on a causal set \cite{ParticlePropagators}.

The ultimate aim would be to develop physically realistic quantum field theories on a causal set. Not only would one expect such field theories to be better defined than their continuum counterparts but one would hope to obtain new predictions for the behaviour of matter which could provide evidence for spacetime discreteness.

A \emph{causal set} (or \emph{causet}) is a locally finite partial order, i.e.
a pair $(\CS,\preceq)$ where $\CS$ is a set and $\preceq$ a relation on $\CS$ which is (i) reflexive ($x\preceq x$); (ii) antisymmetric ($x \preceq y \preceq x \implies  x = y$); (iii) transitive ($x \preceq y \preceq z \implies x \preceq z$); and (iv) locally finite ($\left| \{z' \in \CS | x \preceq z' \preceq y\}\right| < \infty$) for all $x, y, z\in \CS$. Here $\left| A \right|$ denotes the cardinality of a
set $A$. We write $x \prec y$ if $x \preceq y$ and $x \neq y$.

The set $\CS$ represents the set of spacetime events and the partial order
$\preceq$ represents the causal order between pairs of events. If $x \preceq y$
we say ``$x$ is to the causal past of $y$''. The causal relation of a Lorentzian manifold (without closed causal curves) satisfies Conditions (i)-(iii). It is Condition (iv) that enforces spacetime discreteness---each causal interval contains only a finite number of events.

A \emph{sprinkling} is a way to generate a causal set from a $d$-dimensional
Lorentzian manifold $(M,g)$. Points are placed at random within $M$ 
using a Poisson process (with density $\rho$) so the expected number of points in a region of $d$-volume $V$ is $\rho V$.
This generates a causal set whose elements are the sprinkled points and whose partial order
relation is ``read off'' from the manifold's causal relation restricted to the
sprinkled points.

Here we shall only consider causal sets generated by sprinkling into $d$-dimensional Minkowski spacetime, $\mathbb{M}^d$. Such causal sets provide a discretisation of $\mink^d$ which, unlike a regular lattice, is Lorentz invariant \cite[Sec 1.5]{Causet2}.

A \emph{link} between $u, v \in \CS$ (written $u \link v$) is a relation $u \prec v$ such that there exists no $w \in \CS$ with $u \prec w \prec v$.

A \emph{linear extension} of the causal set $(\CS,\preceq)$ is a total order $(\CS,\leq)$ which is
consistent with the partial order. This means that $u \preceq v \implies u
\leq v$ for all $u, v \in \CS$.

Labelling the elements of a finite causal set (with $p$ elements) $v_1,\ldots,v_p$ there are two $p \times p$ \emph{adjacency
matrices}: the \emph{causal matrix} $C$ and \emph{link matrix} $L$ defined by 
\be
C_{xy} := \left\{
\begin{array}{ll} 1 & \textrm{ if } v_x \prec v_y \\ 0 & \textrm{otherwise,}
\end{array} \right. \quad
L_{xy} := \left\{
\begin{array}{ll} 1 & \textrm{ if } v_x \link v_y \\ 0 & \textrm{otherwise.}
\end{array} \right. \ee 

\section{Free Scalar Quantum Field Theory in the continuum}
We briefly review the quantum field theory for a free real scalar field on $\mink^d$ (following the sign conventions of \cite{Bogoliubov:1959nc}). The defining equation for a Green's function $G^{(d)}$ for the Klein-Gordon equation is
 \be \label{eq:KleinGordonGreen} (\Box + m^2)G^{(d)}(x) = \delta^d(x).\ee Here $x = (x^0,\vec{x}) = (x^0,x^i)$ (for $i=1,\ldots,d-1$), $m$ is
the mass of the field, $\delta^d$ is the $d$-dimensional delta function and we choose units with $\hbar = c =1$. The d'Alembertian is $\Box:=
\partial_{x^0}^2 - \partial_{\vec{x}}^2$.

Three important Green's functions (or \emph{propagators}) are the \emph{retarded} $G_R$, \emph{advanced} $G_A$ and \emph{Feynman} $G_F$.
In $\mink^2$ we have \cite{ParticlePropagators}, \cite[p23]{BirrellDavies}: 
\begin{eqnarray}
G_R^{(2)}(x) &:=& \theta(x^0) \theta(s^2) \frac{1}{2} J_0(m s),\\
G_A^{(2)}(x) &:=& G_R^{(2)}(-x),\\
G_F^{(2)}(x) &:=& \frac{1}{4} H_0^{(2)}(m s),
\end{eqnarray}
where $s = \sqrt{(x^0)^2 - (x^1)^2}$ for $(x^0)^2 \geq (x^1)^2$ and
$s=-i\sqrt{(x^1)^2-(x^0)^2}$ for $(x^0)^2 < (x^1)^2$. $J_0$ is a Bessel function of the first kind, $H_0^{(2)}$ is a Hankel
function of the second kind and 
\be \theta(\alpha) = \left\{ \begin{array}{ll} 1
& \textrm{if $\alpha \geq 0$} \\ 0 & \textrm{if $\alpha < 0$}. \end{array}
\right. \ee
In $\mink^4$ we have \cite[Appendix 1]{Bogoliubov:1959nc}: 
\begin{eqnarray}
G_R^{(4)}(x) &:=& \theta(x^0) \theta(s^2) \left(\frac{\delta(s^2)}{2 \pi} - \frac{m}{4 \pi s} J_1(m s)\right), \\
G_A^{(4)}(x) &:=& G_R^{(4)}(-x), \\
G_F^{(4)}(x) &:=& \frac{1}{4 \pi} \delta(s^2) - \frac{m}{8 \pi s} H_1^{(2)}(ms),
\end{eqnarray}
where
 $s = \sqrt{(x^0)^2 - \vec{x}^2}$ for $(x^0)^2 \geq \vec{x}^2$ and
$s=-i\sqrt{\vec{x}^2-(x^0)^2}$ for $(x^0)^2 < \vec{x}^2$. $J_1$ is a Bessel function of the first kind and $H_1^{(2)}$ is a Hankel function of the second kind and $\delta(s^2)$ is the delta function
of $s^2$.

The Pauli-Jordan function is defined as \cite{Bogoliubov:1959nc}
\be \label{eq:PauliJordan} \Delta(x) := G_R(x) - G_A(x).\ee
We drop the dimension-superscripts whenever we refer to arbitrary dimension.

\subsection{Fields in the continuum}

A free real bosonic scalar field is represented by an algebra of field operators
$\p(x)$ (which act on a Fock space $F$) satisfying the following conditions:
\begin{eqnarray}
1. \label{eq:KleinGordon} & & (\Box + m^2)\p(x) = 0,\\
2. \label{eq:Hermitian} & & \p(x) = \p(x)^\dagger,\\
3.\label{eq:Commutator} & & [\p(x),\p(y)] = i \Delta(y-x),
\end{eqnarray}
In addition there exists a Poincar\'e invariant vacuum state $\rv \in F$. With the fields so defined 
the Feynman propagator is given by the vacuum expectation value of the time-ordered product of two field operators:
\be \label{eq:GEDef} G_F(y-x) = i \lv T \p(x) \p(y) \rv. \ee
The time-ordering has time increasing from right to left.

Applying $(\Box_x + m^2)$ to the commutator in \eqref{eq:Commutator} gives 
\be \label{eq:CommutatorKleinGordon} [(\Box_x + m^2)\p(x),\p(y)] = (\Box_x + m^2)i\Delta(y-x) = 0. \ee
That is, even if only \eqref{eq:Hermitian} and \eqref{eq:Commutator} hold, we have that $(\Box_x +
m^2)\p(x)$ commutes with all the $\p(y)$ \footnote{This was emphasised to the author by Johan Noldus.}.

\section{Free Scalar Quantum Field Theory on a causal set}

Let $(\CS, \preceq)$ be a finite causal set with $p$ elements $v_1,\ldots,v_p$ generated by sprinkling (with density $\rho$) into a finite causal interval in $\mink^d$.

For sprinklings into $\mink^2$ the $p \times p$ matrix
\be K_R^{(2)} := a C (I - ab C)^{-1}, \quad a = \frac{1}{2},\;  b=-\frac{m^2}{\rho},\ee
(where $I$ is the identity matrix, $C$ the causal matrix for the causal set and $m$ is the mass of the field) is the appropriate definition of the retarded propagator \cite{ParticlePropagators}.

For sprinklings into $\mink^4$ the $p \times p$ matrix
\be K_R^{(4)} := a L (I - ab L)^{-1}, \quad a = \frac{\sqrt{\rho}}{2\pi\sqrt{6}},\; b=-\frac{m^2}{\rho}, \ee
(where $L$ is the link matrix for the causal set) is the appropriate definition of the retarded propagator.
Both these matrices were defined and studied in detail in \cite{ParticlePropagators}.

Our goal now is to define a $p \times p$ matrix $K_F$ to serve as the Feynman propagator on a causal set. We drop the dimension-superscripts to be able to refer to sprinklings in either $\mink^2$ or $\mink^4$.

If $K_R$ is the retarded propagator it follows that its transpose $K_A := K_R^T$ is the advanced propagator and the real matrix defined by
\be \Delta := K_R - K_A, \ee
is the causal set analogue of the Pauli-Jordan function.

The matrix $i\Delta$ is skew-symmetric and Hermitian. These two properties ensure its
rank is even \cite{Matrices} and its non-zero eigenvalues appear in real positive and negative pairs. In
particular (if its rank is non-zero) there exist non-zero eigenvalues and normalised eigenvectors such
that:
\be i \Delta u_i =\lambda_i u_i, \qquad \qquad i \Delta v_i = -\lambda_i v_i,\ee
(with $\lambda_i > 0$) for $i=1,\ldots,s$ where $2s$ is the rank of $i\Delta$.

These eigenvectors are each defined up to a phase factor and can be chosen such that $u_i = v_i^*$, $u_i^\dagger u_j =
v_i^\dagger v_j = \delta_{ij}$, $u_i^\dagger v_j = 0$ for all $i,j=1,\ldots,s$ (where $z^*$ denotes complex conjugate of $z$ and $u^\dagger:=(u^*)^T$ is the Hermitian conjugate of a column vector $u$).

It's useful to define the Hermitian $p\times p$ matrix \be Q:= \sum_{i=1}^s \lambda_i u_i
u_i^\dagger,\ee
such that $i \Delta = Q - Q^* = Q - Q^T$.

\subsection{Fields on a causal set}

We now define an algebra of field operators $\p_x$ (acting on some Hilbert space $H$) to represent a free real bosonic scalar field on the causal set. 
For each causal set element $v_x$ ($x=1,\ldots p$), we suppose there exists a field operator $\p_x$ such that
\begin{eqnarray}
1. & & \p_x = \p_x^\dagger,\\
2. & &[\p_x,\p_y]= i \Delta_{xy},\\
3. & & \label{eq:ZeroEigenvalueCondition} i\Delta w = 0 \implies \sum_{x'=1}^{p} w_{x'}
\p_{x'} = 0,
\end{eqnarray}
for $x,y = 1,\ldots,p$.
The first two conditions are natural generalisations of the continuum case. The last 
condition ensures that any linear combination of field operators
that commutes with all the field operators must be zero. By \eqref{eq:CommutatorKleinGordon} this is the analogue
of imposing the Klein-Gordon equation on the field operators.

From these field operators we can define new operators
\be \a_i := \sum_{x=1}^{p} (v_i)_x \p_x, \qquad \adag_i := \sum_{x=1}^{p} (u_i)_x \p_x.\ee
for $i=1,\ldots,s$. They satisfy $(\adag_i)^\dagger = \a_i$ and
\begin{eqnarray}
\lbrack \a_i,\a_j\rbrack &=& v_i^T i\Delta v_j = -\lambda_j u_i^\dagger v_j = 0,\\
\lbrack \adag_i, \adag_j\rbrack &=& u_i^T i\Delta u_j = \lambda_j v_i^\dagger u_j = 0,\\
\lbrack \a_i,\adag_j\rbrack &=& v_i^T i\Delta u_j = \lambda_j u_i^\dagger u_j = \lambda_j\delta_{ij},
\end{eqnarray}
and can be interpreted as (unnormalised) creation and annihilation operators.

The transformation can be inverted to give
\be \p_x = \sum_{i=1}^s (u_i)_x \a_i + (v_i)_x \adag_i, \ee
(here it is important we imposed \eqref{eq:ZeroEigenvalueCondition}).

We now define a vacuum state vector $\rv \in H$ by the conditions that $\a_i \rv = 0$ for all
$i=1,\ldots,s$ and $\lv0\rangle = 1$. This allows us to recognise that $H$ is the Fock space spanned by basis vectors
$(\adag_1)^{n_1}(\adag_2)^{n_2}\cdots (\adag_s)^{n_s} \rv$ for integers $n_i \geq0$, $i=1,\ldots,s$.

The two-point function can be evaluated as
\begin{eqnarray}
\lv \p_x \p_y \rv &=& \sum_{i=1}^s \sum_{j=1}^s (u_i)_x (v_j)_y \lv  \a_i
\adag_j \rv \\ \nonumber &=& \sum_{i=1}^s \sum_{j=1}^s (u_i)_x (v_j)_y
\lambda_j \delta_{ij} = Q_{xy}.\end{eqnarray}

We can now define the Feynman propagator by analogy with \eqref{eq:GEDef}. With time increasing from right to left we have
\be \label{eq:KFDef} (K_F)_{xy}:= i \lv T \p_x \p_y \rv := i\left(\bar{A}_{xy} Q_{yx} + \bar{A}_{yx}
Q_{xy} + \delta_{xy} Q_{xy}\right),\ee
where $\Ale$ is the causal matrix for a linear extension of the causal set and $\delta_{xy}$ is the Kronecker delta.
In general there are multiple different linear extensions which assign an
arbitrary order to pairs of unrelated elements. This arbitrariness does not affect 
$K_F$ because the field operators for unrelated elements commute.

Observing that $\Ale_{xy} (i\Delta_{xy}) = (iK_R)_{xy}$ we have
\be \Ale_{xy} Q_{yx} = \Ale_{xy}(Q_{xy}-i \Delta_{xy}) = \Ale_{xy} Q_{xy} -i (K_R)_{xy}.\ee
Substituting this into \eqref{eq:KFDef} gives an alternative form
\be \label{eq:KFAlt} K_F = K_R + i Q,\ee
since $\Ale_{xy} + \Ale_{yx} + \delta_{xy} = 1$ for all $x, y=1,\ldots,p$.

Since $i \Delta = Q - Q^*$ the imaginary part of $Q$ is $\Im(Q) = \Delta/2$. Combining this with \eqref{eq:KFAlt} and looking at the real and imaginary parts of $K_F$ gives
\begin{eqnarray}
\label{eq:KFReal} \Re(K_F) &=& K_R - \frac{\Delta}{2} = \frac{K_R + K_A}{2},\\
\label{eq:KFImag} \Im(K_F) &=& \Re(Q).
\end{eqnarray}

\section{Comparison with the Continuum}

The causal set propagators depend on the particular random causal set that is sprinkled. By calculating their average value for different sprinklings we can compare the causal set and continuum propagators. To do this, first fix a finite causal interval $\cint \subset \mink^d$. Pick two points $\X, \Y \in \cint$. Sprinkle a finite causal set into $\cint$ with density $\rho$. Almost surely this will not contain $\X$ and $\Y$ so add $\X$ and $\Y$ to it to obtain a finite causal set $(\CS,\preceq)$. For definiteness label the causal set element $\X$ as $v_1$ and $\Y$ as $v_2$.

We now calculate $K_R$ and $K_F$ for $(\CS,\preceq)$ and look at $(K_R)_{12}$ and $(K_F)_{12}$, i.e. the propagator values for the pair $(\X, \Y)$. Let $\mathbb{E}(K_R^{(d)}|\X, \Y, \mink^d, \rho)$ denote the expected value of $(K_R^{(d)})_{1 2}$ (and $\mathbb{E}(K_F|\X, \Y, \mink^d, \rho)$ the expected value of $(K_F)_{1 2}$) averaged over all causal sets sprinkled into $\cint \subset \mink^d$ (with $\X$ and $\Y$ added in the manner described and for a fixed density $\rho$). It was shown in \cite{ParticlePropagators} that
\begin{eqnarray}
\label{eq:1+1expected} \mathbb{E}(K_R^{(2)}|\X, \Y,\mathbb{M}^2,\rho) &=& G_R^{(2)}(\Y-\X),\\ 
\label{eq:3+1expected} \lim_{\rho\to\infty}\mathbb{E}(K_R^{(4)}|\X,\Y,\mathbb{M}^4,\rho) &=& G_R^{(4)}(\Y-\X).
\end{eqnarray}
Using these and \eqref{eq:KFReal} we have
\begin{eqnarray}
\label{eq:KF1+1} \Re(\mathbb{E}(K_F|\X,\Y,\mathbb{M}^2,\rho)) &=& \Re(G_F^{(2)}(\Y-\X)),\\
\label{eq:KF3+1} \lim_{\rho\to\infty} \Re(\mathbb{E}(K_F|\X,\Y,\mathbb{M}^4,\rho))&=& \Re(G_F^{(4)}(\Y-\X)).
\end{eqnarray}
That is, the real part of the expected value of $K_F$ is correct for $\mink^2$ and correct in the infinite density limit for $\mink^4$ \footnote{$\Re(G_F(x)) = (G_R(x) + G_A(x))/2$.}. We can compare the imaginary parts of $K_F$ and $G_F$ through numerical simulations.

By using a computer to perform sprinkling into finite regions of 1+1 and 3+1 Minkowski spacetime we can compute $K_R, K_A, i\Delta, Q$ and $K_F$ for a range of sprinkling densities and field masses. Plotting the real and imaginary parts of $K_F$ against the absolute value of the proper time for causally related and spacelike separated pairs of sprinkled points allows us to see if $K_F$ agrees with $G_F$.

\begin{figure}
\begin{center}
\includegraphics[width=0.5\textwidth]{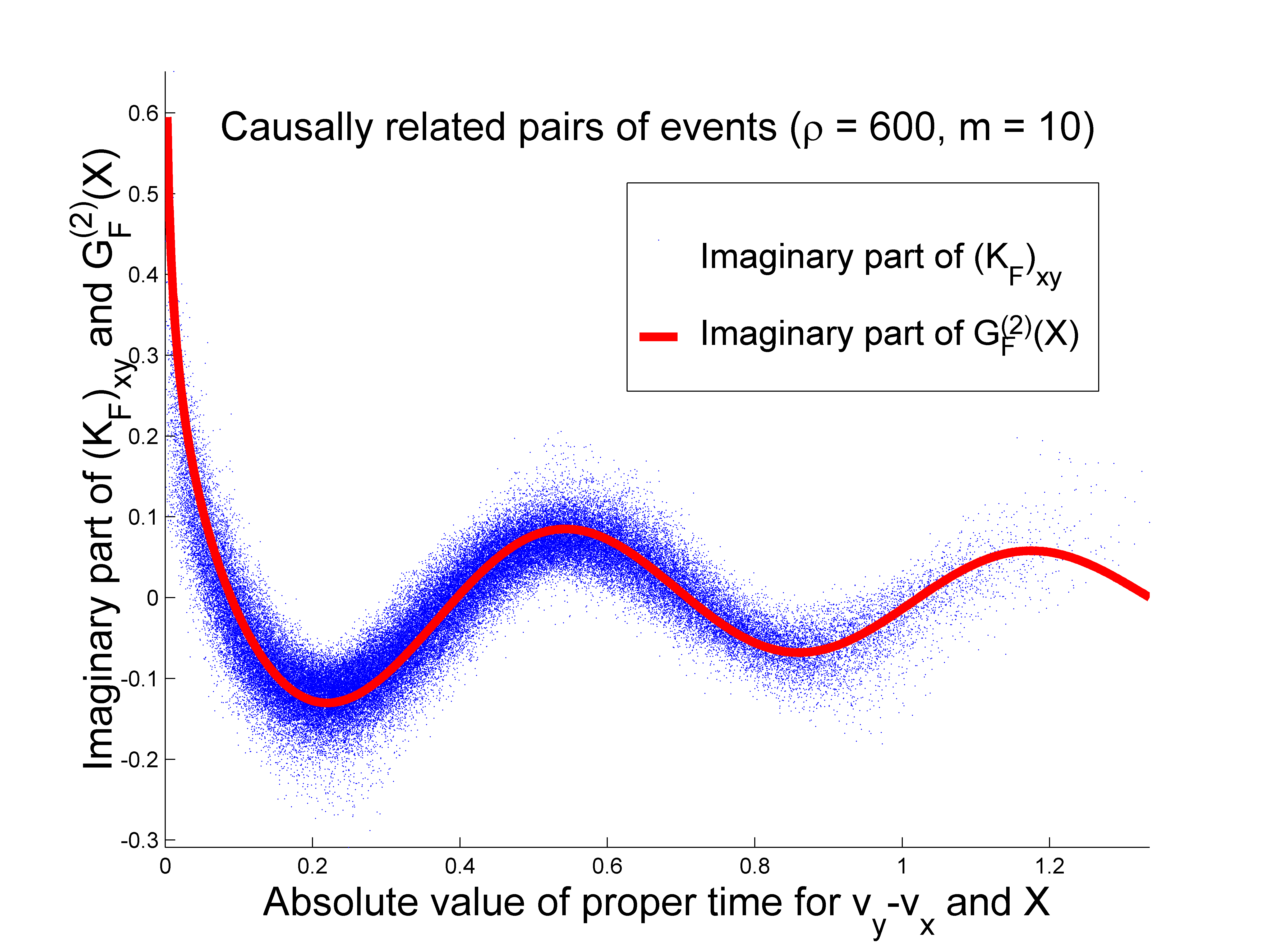}
\includegraphics[width=0.5\textwidth]{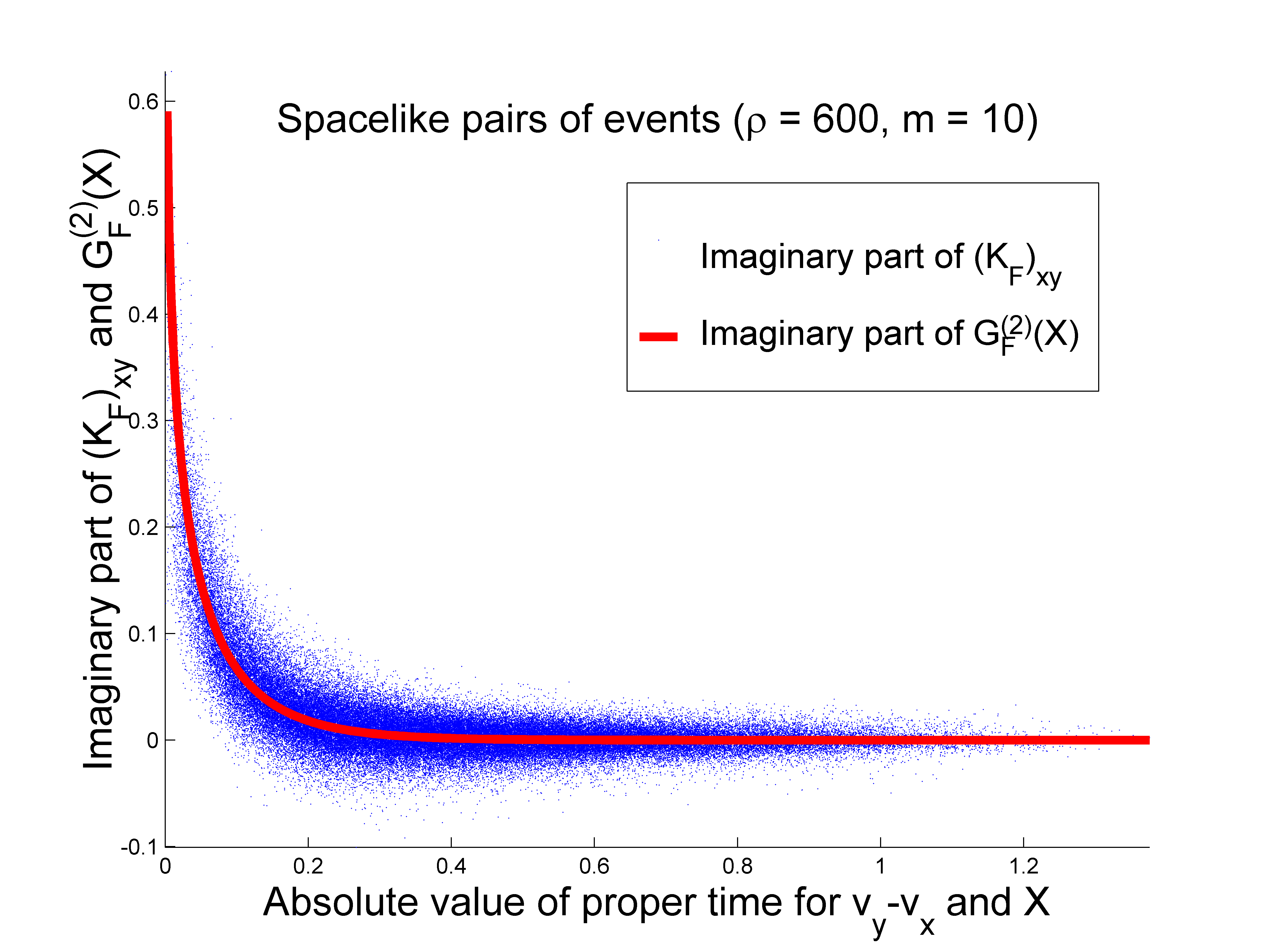}
\caption{\label{fig:Plots} Plots of $\Im((K_F)_{xy})$ for pairs of sprinkled points $v_x$ and $v_y$ in a \emph{single} 600 element causal set generated by sprinkling into a unit-volume causal interval in $\mink^2$ and $\Im(G_F^{(2)}(X))$ for $X \in \mink^2$.}
\end{center}
\end{figure}

Simulations using standard linear algebra software on a desktop computer give clear results for the 1+1 dimensional case for causal sets with as few as 600 elements (see Fig.\ \ref{fig:Plots}). The agreement is very good provided $0 \ll m \ll \sqrt{\rho}$. There is
disagreement between the imaginary parts of $K_F$ and $G_F^{(2)}$ as we take the field mass to zero but this seems to be related to the lack of a massless limit of $\Im(G_F^{(2)})$. There are also small deviations due to the calculation being sprinkled into a \emph{finite} region of Minkowski spacetime. These ``edge effects'' get smaller if the results are only plotted for pairs of points away from the edges of the region.

In 3+1 dimensions the simulations are less clear. We expect good agreement with the continuum for large $\rho$ (since the infinite density limit is needed in \eqref{eq:3+1expected}) but larger sprinklings are needed to investigate this.

\section{Discussion}

The formalism provokes interesting questions about the eigenvectors $u_i, v_i$. They seem to be related to linear combinations of positive and negative frequency plane wave solutions of the Klein-Gordon equation. Plotting the eigenvector values as a function of position for causal sets sprinkled into $\mink^2$ shows relatively smooth oscillatory behaviour for the eigenvectors associated with large eigenvalues. They show some dependence on the shape of the region into which the points are placed but this dependence is lost when $\lambda_i u_i u_i^\dagger$ is summed in the $Q$ and $K_F$ matrices (details to appear elsewhere).

Extending $K_F$ to infinite causal sets as well as further investigating the behaviour of $K_F$ for sprinklings into $\mink^4$ as well as other conformally flat or curved spacetimes would be of significant interest. Investigating the variance of $K_F$ for different sprinkling densities is also important. The propagator $K_R$ (and therefore $K_F$) can be extended to non-sprinkled causal sets \cite[Sec 4.2]{ParticlePropagators}. This ensures both $K_R$ and $K_F$ are relevant to a fundamental theory based only on causal sets with no reference to Lorentzian manifolds or sprinklings.

The work presented here opens the door to interacting scalar field theories, scalar field phenomenology, the evaluation of Feynman diagrams, S-matrices and scattering amplitudes on a causal set. Extending the quantum field theory to include interacting spinor and vector fields remains to be done---advances here could lead to physical predictions for matter on a causal set.

The author would like to thank Fay Dowker for ongoing support and encouragement. Also Rafael Sorkin, Johan Noldus for helpful early discussions.
This work was supported by the STFC and in part by Perimeter Institute for Theoretical Physics.

\end{document}